\documentclass[10pt,twocolumn,twoside]{opticajnl}
\journal{opticajournal} 

\setboolean{shortarticle}{false}

\usepackage{lineno}

\usepackage{color}
\usepackage{graphicx}
\usepackage{bm}
\usepackage{braket}

\usepackage{tabularx}
\usepackage{booktabs} 

\title{Construction of Boolean Logic Gates Using QFT-Based Adder Architecture}

\author[1]{Ayda Kaltehei}
\author[2]{Murat Kurt}
\author[1]{Azmi Gen\c{c}ten}
\author[2,*]{Sel\c{c}uk \c{C}akmak}

\affil[1]{Department of Physics, Ondokuz May{\i}s University, 55139 Samsun, T\"{u}rkiye.}
\affil[2]{Department of Software Engineering, Samsun University, 55420 Samsun, T\"{u}rkiye.}

\affil[*]{selcuk.cakmak@samsun.edu.tr}

\begin{abstract}
In this study, we construct the quantum reversible counterparts of the logical AND, OR, XOR, NOR, and NAND gates. We utilize a quantum Fourier transform (QFT)-based adder circuit that replicates the functionality of a digital half-adder, which computes the sum and carry of two input bits using XOR and AND gates, respectively. To realize different logic gate operations, we apply pre- and post-processing to the QFT-adder using quantum gates, leveraging Boolean algebra properties to enable conversions between various logical functions. Although the number of elementary quantum logic gates increases for a small number of inputs-making the approach appear inefficient at first glance-the overall required qubit count is reduced compared to non-QFT-based designs as the number of inputs increases.
\end{abstract}

\setboolean{displaycopyright}{false} 

\begin{document}

\maketitle

\section{\label{sec:intro} Introduction}
Computers perform highly complex operations in information processing processes. These operations are, in fact, carried out through a series of calculations and logic. In this process, classical computers rely on basic components called logic gates. Logic gates process digital signals and perform various operations on data; these operations enable the computer to execute the defined algorithms. The arithmetic operations can be also performed using elementary logic operators such as AND, OR, and NOT, to carry out complex data processing tasks. These small yet powerful building blocks are the hidden heroes behind the ability of computers to process millions of data in seconds. However, classical computers are limited to these basic logic gates and digital processing. Although they can process millions of data in seconds, they still have significant limitations when it comes to complex calculations and certain problems. At this point, quantum computers come into play. Quantum computers rely on a completely different approach to information processing, one that promises to overcome the limitations of traditional computers. By going beyond classical logic gates and utilizing principles of quantum mechanics, such as quantum superposition and quantum entanglement, they have the capacity to process much larger and more complex data sets~\cite{Bennet2000} . These new-generation computers have the potential to revolutionize fields like encryption, artificial intelligence, and big data analysis~\cite{Chae2024}. At the present stage, although there are some physical limitations in processing big data in quantum information processing, it is believed that these limitations will be overcome in the future.

In quantum computing, information is encoded into the quantum states of quantum mechanical particles and processed according to the laws of quantum mechanics. The fundamental unit in which information is encoded is called a qubit. Due to the properties of superposition, entanglement, and no-cloning, qubits have both higher capacity and greater security compared to classical bits. These properties contribute to what is known as quantum supremacy~\cite{Harrow17,Boixo18,Arute19}.

Quantum algorithms use quantum phenomena like superposition and entanglement to perform operations in parallel. Quantum circuits that execute quantum algorithms use quantum bits (qubits). Quantum circuits process information by applying quantum gates to qubits. The key difference between quantum gates and classical logic gates is that quantum gates are reversible. This property ensures that no data is lost~\cite{Nielsen12}. Parallel processing in quantum information processing occurs because all the information is available simultaneously. This means that one or more inputs are put into a superposition state. If there is a single input, this is achieved using the Hadamard gate; if there are multiple inputs, it is also achieved using the quantum Fourier transform sub-circuit. 

The quantum Fourier transform (QFT) is a powerful algorithm that plays a crucial role in quantum computing and is used to solve various problems~\cite{Shor97,Grover96}. QFT is the quantum counterpart of the classical Fourier transform and, by leveraging quantum properties like superposition and entanglement, it can perform certain computations significantly faster than classical algorithms~\cite{Camps21, Draper20, Ruiz17, Sahin20, Cakmak24}.

This study explores the construction of the quantum counterparts of Boolean logic gates, namely AND, NAND, OR, NOR, and XOR, using a quantum Fourier transform-based adder circuit, which serves as a fundamental building block for constructing more complex digital circuits. This approach reduces the required quantum resource (number of the qubits) for logic gates with large input sizes and also offers the advantage of reduced noise arising from interference between quantum channels.

\section{\label{sec:theory} Theory}
In this section, the quantum Fourier transform (QFT) and the QFT-based adder will be introduced. The required quantum logic gates and the functionality of each gate used in these circuits will be explained in detail.
\subsection{\label{bit}Quantum bit}
The qubit, the basic unit in quantum computing, is a two-dimensional vector space over the complex numbers. 
The basis states for a qubit are represented as $\ket{0}$ and $\ket{1}$. A qubit is not limited to being in only basis state. It can exist in an arbitrary quantum state $\ket{\psi}$, which is a superposition expressed as;
\begin{equation}\label{qubit}
\ket{\psi}=\alpha\ket{0}+\beta\ket{1}
\end{equation}
here, $\alpha$ and $\beta$ are complex numbers satisfying the normalization coefficients, 
\begin{equation}
|\alpha|^2 + |\beta|^2 = 1
\end{equation}
because a qubit can exist as both $0$ and $1$ simultaneously, with associated probabilities, one can conceptualize it as existing in multiple parallel universes.  For instance, a qubit could be seen as splitting, with different probabilities, into two parallel universes and then potentially recombining into a photon representing either $0$ or $1$. This parallelism allows for faster calculations compared to classical computing~\cite{Abdelgaber2020}.

\subsection{\label{qft}Quantum Fourier Transform}
In quantum computing, the QFT places inputs into a superposition state, leading to the parallel processing of multiple inputs~\cite{Camps21}. The QFT consists of Hadamard, Controlled Phase Shift, and SWAP gates. It is found in the structure of leading algorithms in quantum information processing. The inverse quantum Fourier transform (IQFT) undoes the QFT, converting a state that has been transformed into the Fourier basis back into its original computational basis representation~\cite{Jozsa98,Barenco96,Cao11}. The generic expression of the QFT operation can be given as,
\begin{equation}\label{QFT}
\mathrm{QFT}\ket{a} = \frac{1} {\sqrt{2^n}}\sum_{k=0}^{{2^n}-1}e^{2\pi i.ak/{2^n}}\ket{k} 
\end{equation}
Here, $\ket{a}$ represents n-qubit quantum state, and $\ket{k}$ represents the superposition state to be phase encoded in Fourier space ~\cite{Ruiz17}. Both $\ket{a}$ and $\ket{k}$ are represented by $2^n\times1$ dimensional matrices. The QFT itself is a $2^n\times2^n$ dimensional square matrix.

Similarly, the mathematical equality of the IQFT operator is shown as follow;
\begin{equation}\label{IQFT}
\mathrm{IQFT}\ket{k} = \frac{1}{\sqrt{2^n}}\sum_{a=0}^{{2^n}-1}e^{-2\pi i.ak/{2^n}}\ket{a}
\end{equation}

In this study we use, QFT operator in $2^2\times2^2$ dimension, where the quantum circuit to apply QFT operation is constructed as; 

\begin{figure}[!htb]\centering
\includegraphics[width=6.5cm]{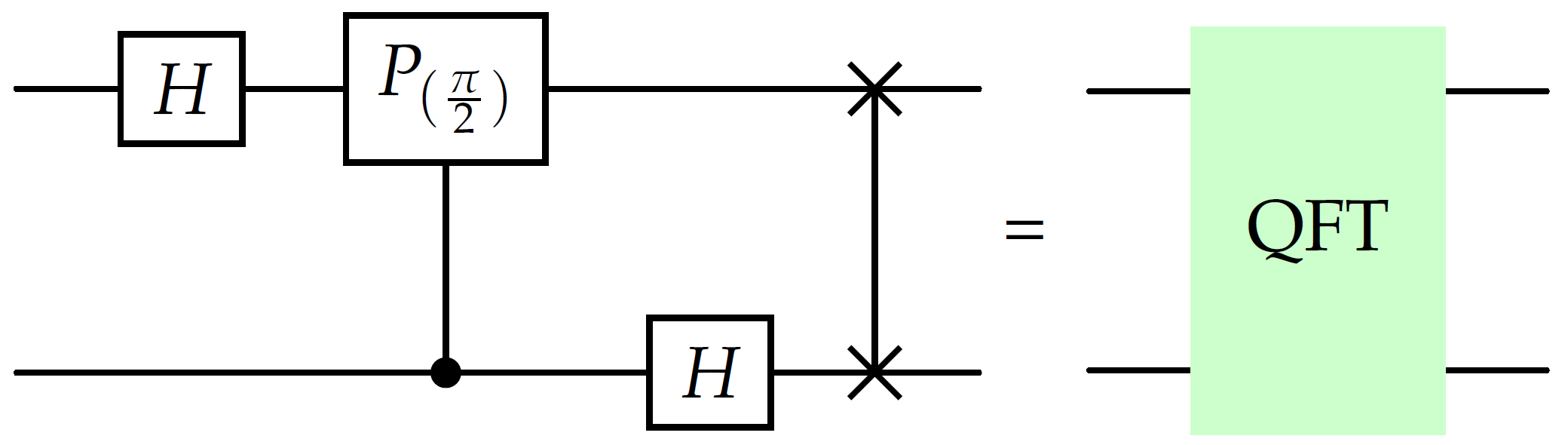}
\caption{\label{qft_two} Two-qubit quantum Fourier transform circuit.}
\end{figure}

\begin{figure}[!htb]\centering
\includegraphics[width=6.5cm]{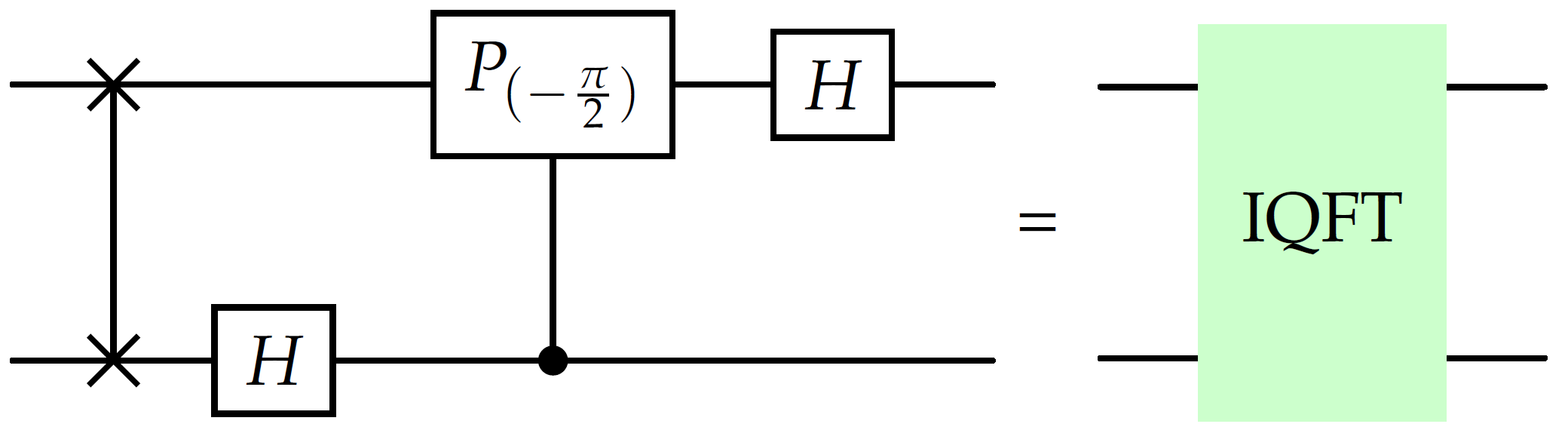}
\caption{\label{iqft_two} Two-qubit inverse quantum Fourier transform circuit.}
\end{figure}

The Figs.~\ref{qft_two} and~\ref{iqft_two} show two-input QFT and IQFT circuits, respectively. In Fig.~\ref{qft_two}, first, the Hadamard gate acts on a single qubit, creating a superposition state. The general matrix form of the Hadamard gate for an $n$-qubit system is expressed as follows:
\begin{equation}
H^{\otimes n} = \frac{1}{\sqrt{2^n}} \sum_{x, y \in \{0,1\}^n} (-1)^{x \cdot y} \ket{x}\bra{y}
\label{eq:hadamard_tensor}
\end{equation}
where $x \cdot y = x_0 y_0 \oplus x_1 y_1 \oplus \cdots \oplus x_{n-1} y_{n-1}$ denotes the bitwise dot product modulo 2. $\ket{x}$ and $\bra{y}$ are computational basis kets and bras, respectively.
The Hadamard gate $H$ for a single qubit is defined by the matrix:
\begin{equation}
H = \frac{1}{\sqrt{2}}
\begin{bmatrix}
1 & 1 \\
1 & -1
\end{bmatrix}
\label{eq:hadamard}
\end{equation}

Second, the phase shift gate, like Hadamard, acts on a single qubit. If the input is $\ket{0}$, it does not make any change, but it adds a phase if the input is $\ket{1}$. The matrix representation of the phase shift gate for qubits is shown below.
\begin{equation}
    P(\theta_k) =\begin{bmatrix}
        1 & 0 \\
        0 & e^{i\theta_k} \\
    \end{bmatrix}
\end{equation}here, $\theta_k = \frac{2\pi}{2^k}$. However, the QFT operator includes controlled phase shift gates. If the control input is $\ket{1}$, the phase shift gate in equation 10 is applied to the target qubit. The following equation provides the equation that finds the matrix representation of the phase shift gate for qubits \cite{Pavlidis17,Kurt24}.
\begin{equation}
    CP(\theta_k)= \sum_{j=0}^{1} \sum_{m=0}^{1} {e^{\frac{i2\pi}{2^k}jm}} \ket{j}\bra{j} \otimes \ket{m}\bra{m}
\end{equation}

Lastly, SWAP gate is applied to order of the outputs in Fourier phase space.

\subsection{\label{sec:qft-adder}Quantum Fourier transform-based adder}

\begin{figure}[!htb]\centering
\includegraphics[width=7.5cm]{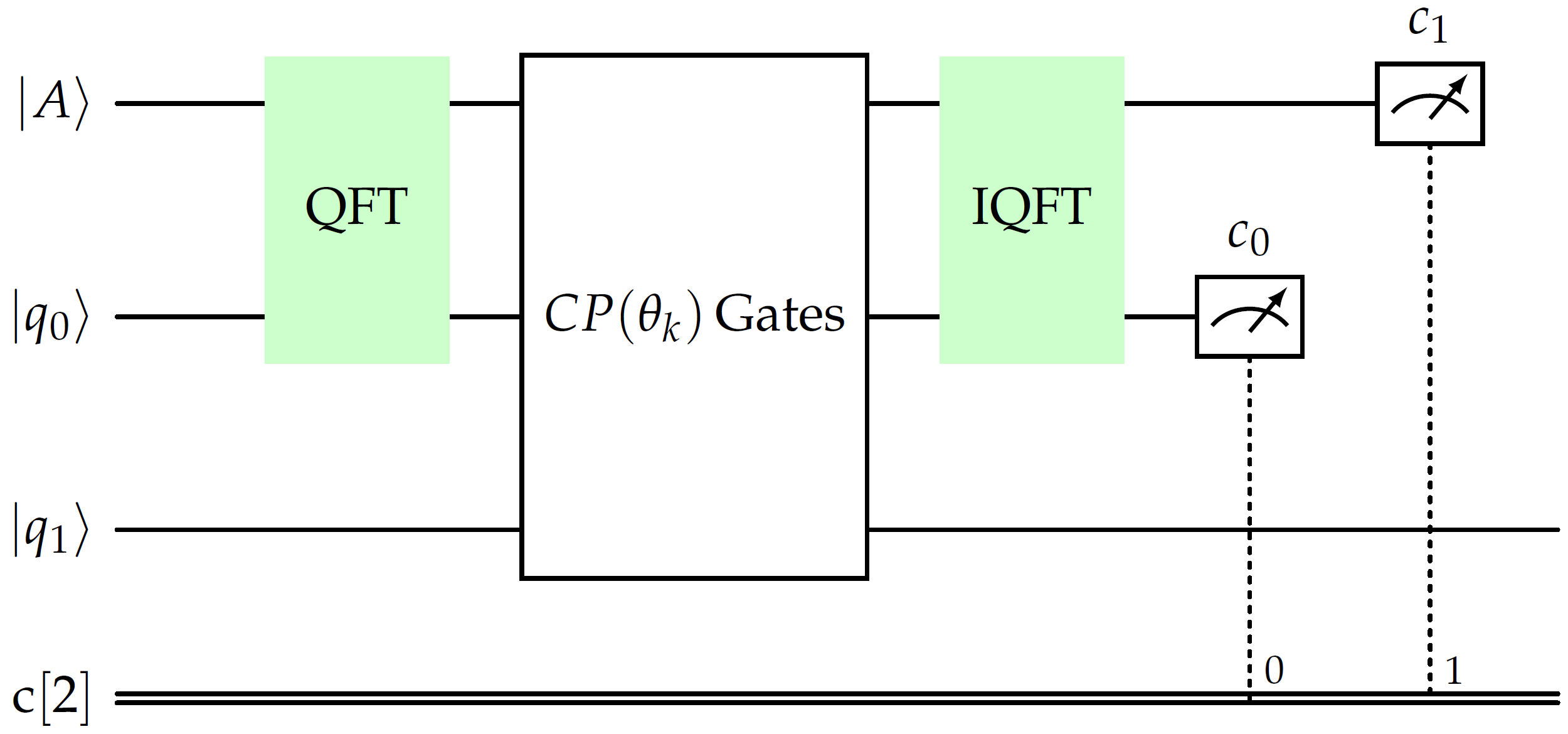}
\caption{\label{2bit_adder} The quantum circuit diagram of the one-bit two input QFT-based adder. The carry bit and sum bit, which hold the result of the addition operation, are stored in $c_1$ and $c_0$, respectively.}
\end{figure}

\begin{figure}[!htb]\centering
\includegraphics[width=5.5cm]{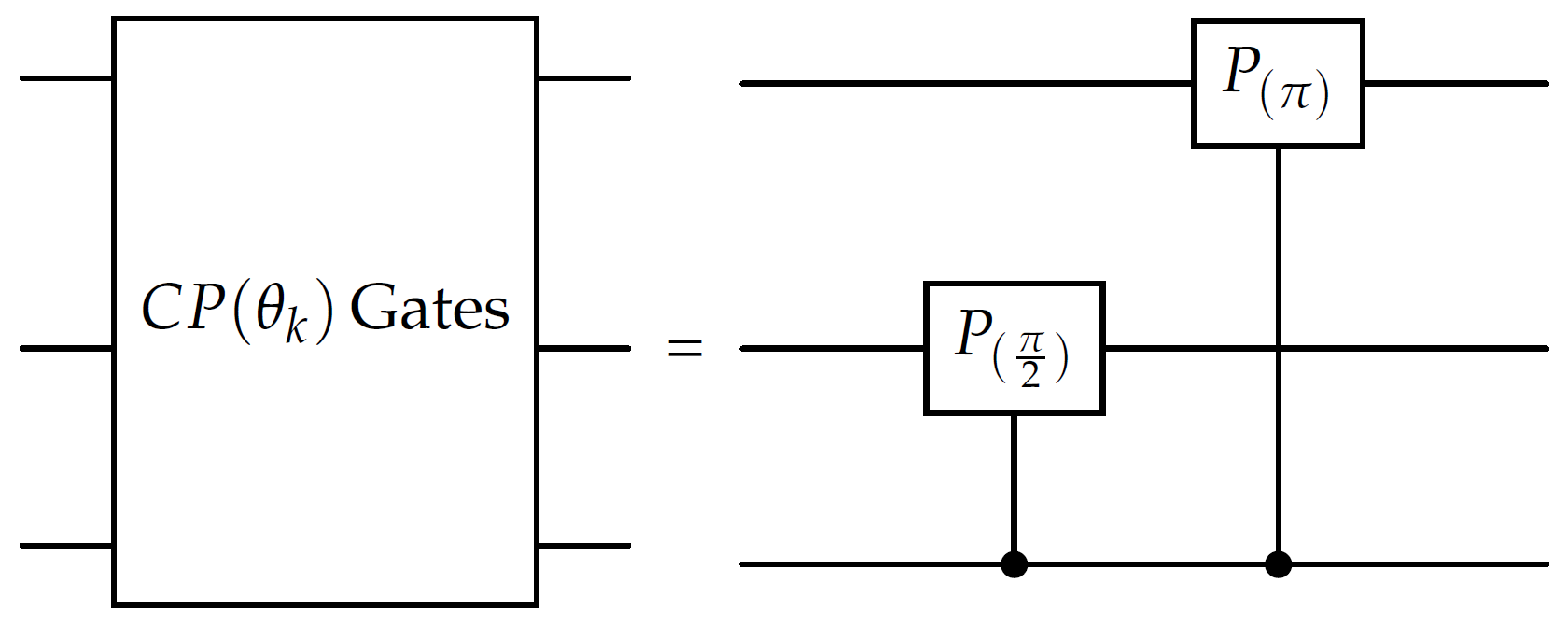}
\caption{\label{adder_part} The $CP(\theta_k)$ gates for one bit two-input QFT-adder.}
\end{figure}

In this research, we use QFT-based adder to construct Boolean logic gates. The generic structure of the QFT-based adder can be shown as in Fig.~\ref{2bit_adder}. The two input qubits, $\ket{q_0}$ and $\ket{q_1}$, hold the numerical information on which the addition operation is performed, while $\ket{A}$ is an ancillary qubit initially set to $\ket{0}$. The measurement is applied on the $\ket{A}$ qubit gives the Carry-bit encoded in $c_1$ bit, while the measurement of the $\ket{q_0}$ provides the Sum-bit encoded in $c_0$. The QFT-based adder operates in the following steps: First, after applying the QFT to the first two inputs, the system transitions into the Fourier basis. So, the first two quantum channels are in a superposition state, with a phase factor applied to each qubit. Second, a set of conditional phase shift gates acts on all three channels, as shown in Fig.~\ref{adder_part}. At this point, the phase states originating from $\ket{q_1}$ are added to the results obtained from the first two channels. This step is where the addition is realized. In the final step, the IQFT is applied to transform the system back to the computational basis.

\maketitle
\section{\label{sec:Results} Results and Discussion}
\begin{figure}[!htb]\centering
\includegraphics[width=5.5cm]{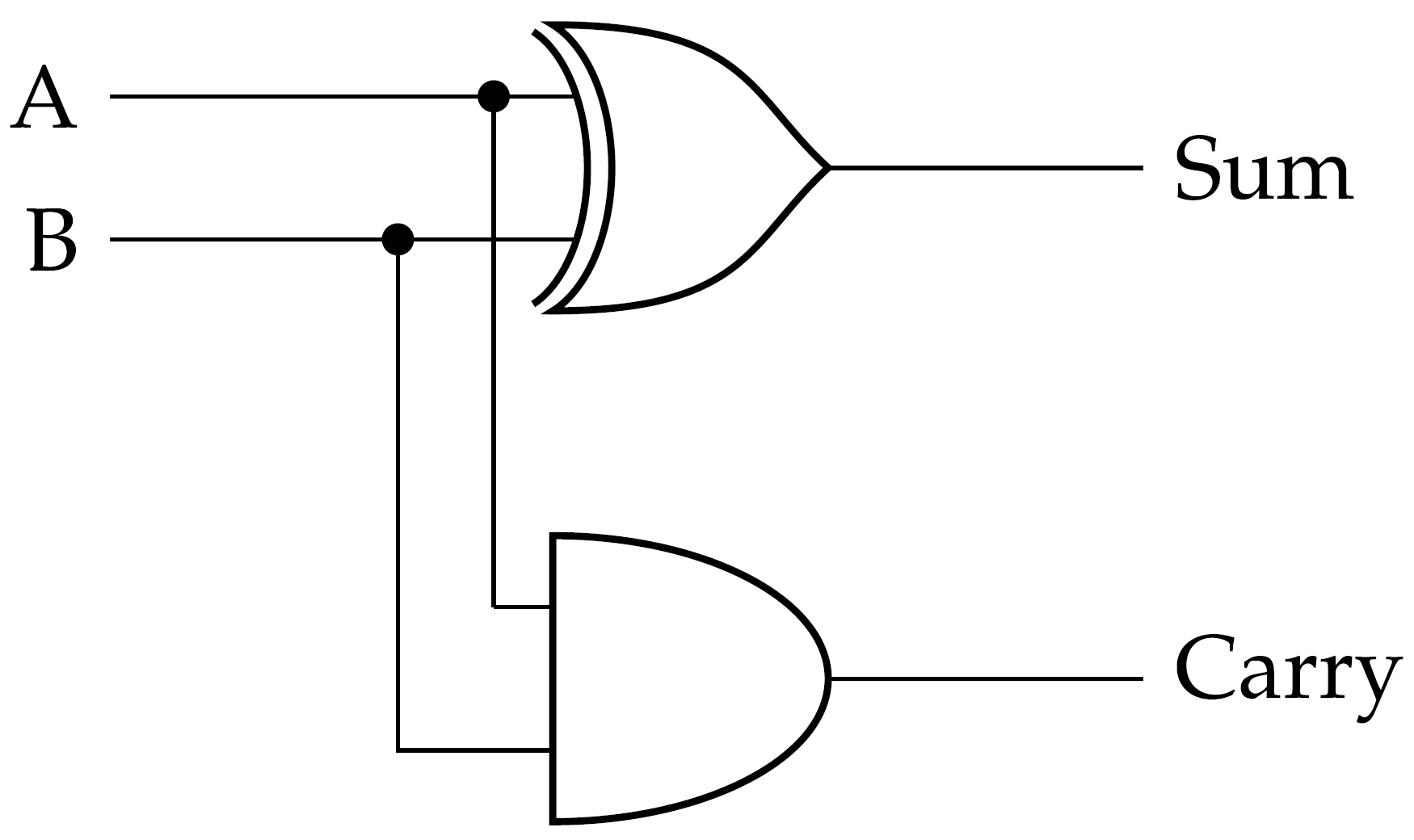}
\caption{\label{half-adder} The classical half-adder, where A and B represent the two input bits, and the Sum and Carry outputs hold the result of the addition operation.}
\end{figure}

In this section, we construct the AND, NAND, OR, NOR, and XOR logic gates using a QFT-based adder, all of which are reversible—unlike their classical counterparts. To implement these logic functions with quantum gates, we refer to the classical half-adder circuit shown in Fig.~\ref{half-adder}, which includes an XOR gate for the sum output and an AND gate for the carry output of the adder.

\begin{figure}[!htb]\centering
\includegraphics[width=8.5cm]{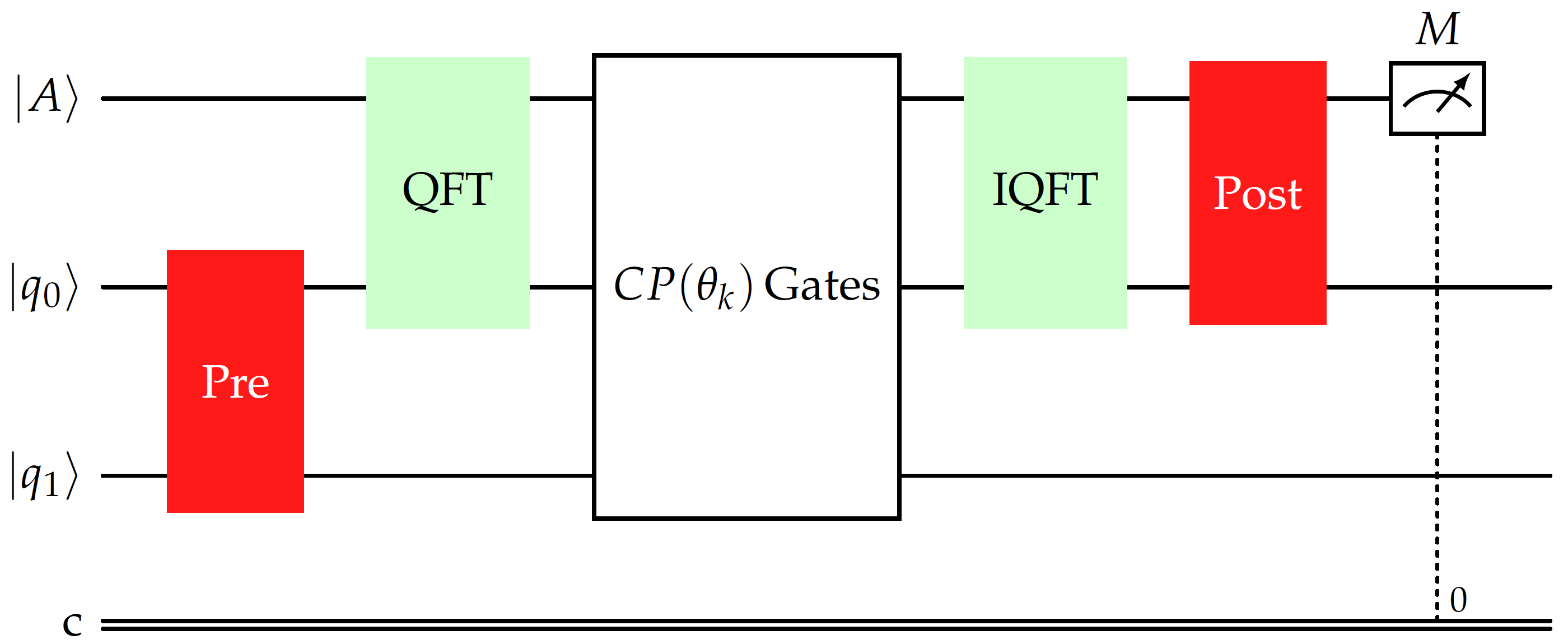}
\caption{\label{PrePostQFTadder} The generic circuit design of the quantum counterparts of logic gates is based on the QFT-adder. The two input qubits, $\ket{q_0}$ and $\ket{q_1}$, each represent one bit of information on which the logical operation is performed. The result is obtained by measuring the ancillary qubit $\ket{A}$, with the outcome stored in the classical bit $c$.}
\end{figure}

The generic form of the quantum circuit for the QFT-based adder operation is presented in Sec.~\ref{sec:qft-adder}. We apply pre- and post-processing operations to the QFT-based adder, as shown in Fig.~\ref{PrePostQFTadder}, using quantum gates to perform Boolean algebra properties, such as De Morgan's laws. Thus, we obtain the Boolean logic gates by utilizing a QFT-based adder circuit that performs the function of a classical half-adder. Furthermore, we evaluate the accuracy of each logic gate using the IBM quantum composer~\cite{IBMqc}.

Quantum equivalents of classical logic gates have also been explored through various circuit implementations that do not rely on the quantum Fourier transform~\cite{Renaud11,Paul23,Miller11}. For example, a two-input NAND gate can be implemented in quantum computing using the Toffoli (CCNOT) gate—which requires three qubits. However, this approach becomes increasingly impractical as the number of inputs grows. Specifically, constructing an $N$-input NAND gate in this manner demands $2N - 1$ qubits, resulting in high resource consumption. By extracting the carry-out bit from a one-bit ($n=1$) $N$-input QFT-based adder circuit, the NAND operation can be efficiently achieved. This method reduces the total qubit requirement to $N + \log_2 N$, providing a scalable and resource-efficient solution for implementing multi-input NAND logic in quantum circuits~\cite{Cakmak24}.

\subsection{\label{AND} Implementation of the logic AND gate}
\begin{figure}[!htb]\centering
\includegraphics[width=8.5cm]{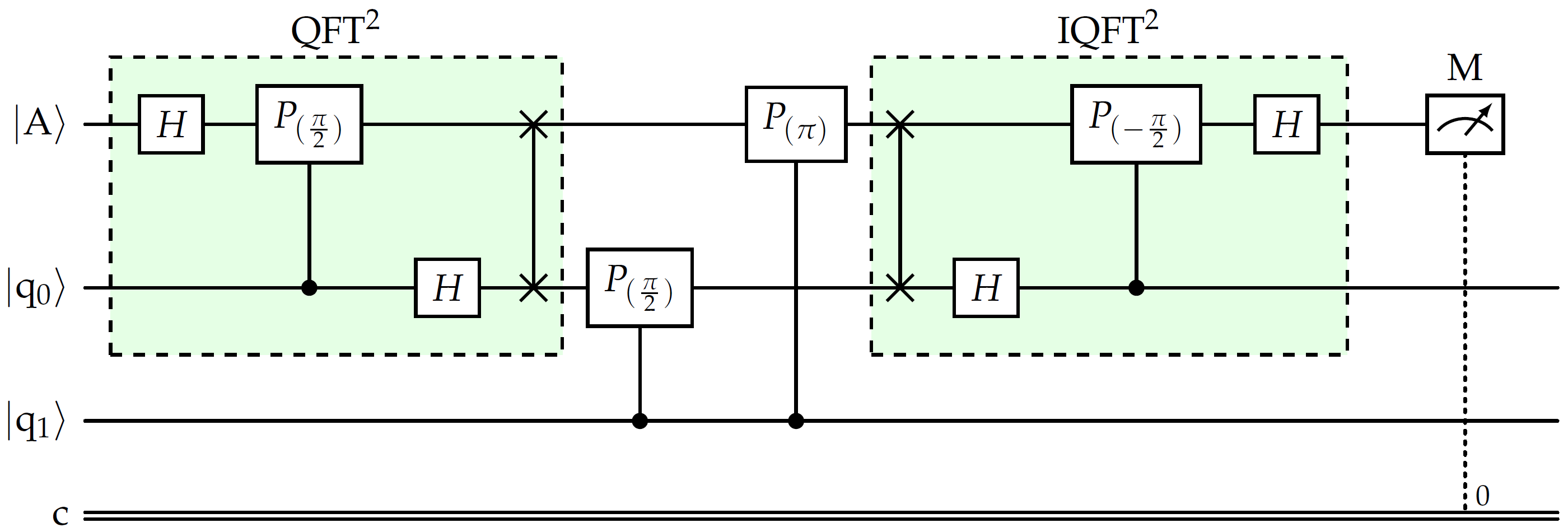}
\caption{\label{and_gate} The QFT-adder based two-input AND gate.}
\end{figure}

To design a two-input logic AND gate, we employ a QFT-based quantum adder circuit. This circuit consists of three inputs: one ancillary qubit and two input (data) qubits, represented by $\ket{A}$, $\ket{q_0}$, and $\ket{q_1}$, respectively. As shown in the one-bit, two-input QFT-based adder circuit in Fig.~\ref{and_gate}, the measurement of the ancillary qubit yields the carry bit, which corresponds to the AND gate output in a digital half-adder. No pre- or post-processing quantum gates are applied to the QFT-based adder circuit. The measurement process maps the result of the AND gate, to a classical bit, $\mathrm{c}=q_0$ AND $q_1$).

The QFT-adder based AND circuit was tested on IBM quantum composer for all input states, $\ket{00}$,$\ket{01}$, $\ket{10}$ and $\ket{11}$ (in the $\ket{q_0q_1}$ form), and the results were verified as given in Table~\ref{tab:and_gate_measurements}
\begin{table}[ht]
\caption{\label{tab:and_gate_measurements}Truth table of the QFT-based 2-input AND-gate.}
\noindent
\begin{tabularx}{\columnwidth}{@{}XXX@{}} 
\toprule
$\ket{q_0}$ & $\ket{q_1}$ & $c$ \\
\midrule
$\ket{0}$ & $\ket{0}$ & \text{0} \\
$\ket{0}$ & $\ket{1}$ & \text{0} \\
$\ket{1}$ & $\ket{0}$ & \text{0} \\
$\ket{1}$ & $\ket{1}$ & \text{1} \\
\bottomrule
\end{tabularx}
\end{table}

\subsection{\label{NAND}Implementation of the logic NAND gate}
\begin{figure}[!htb]\centering
\includegraphics[width=8.5cm]{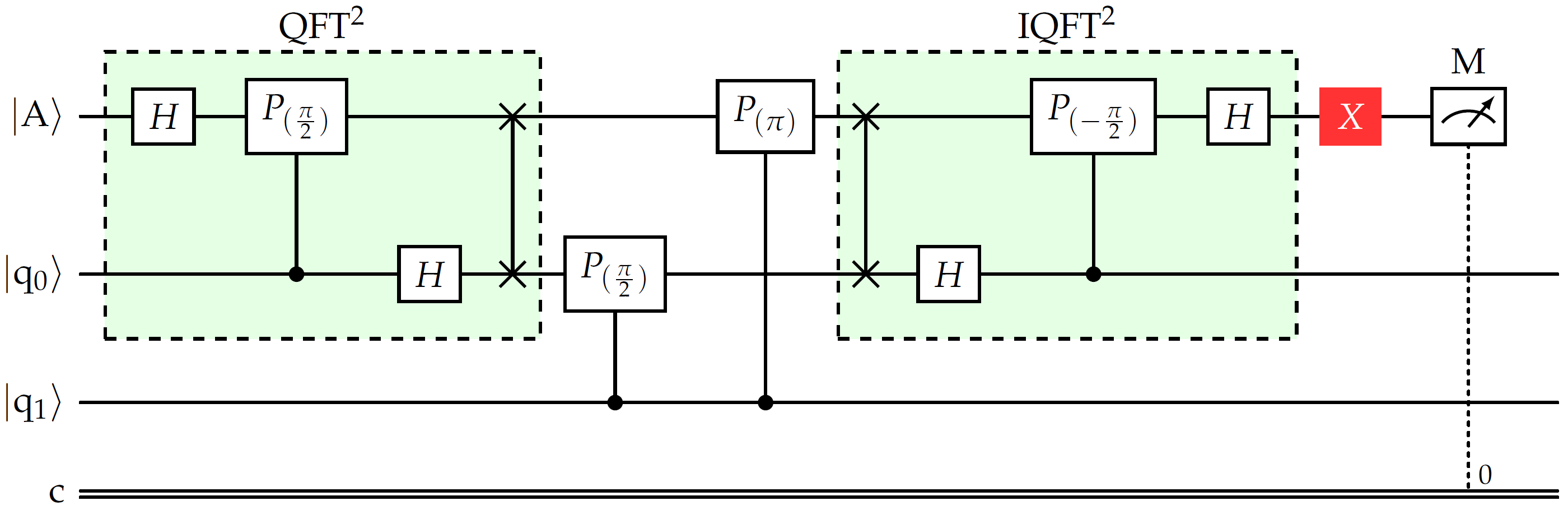}
\caption{\label{nand_gate} The QFT-adder based two-input NAND gate.}
\end{figure}

A two-input logic NAND (NOT-AND) gate is constructed via applying the X-gate, which represent classical NOT gate, to QFT-based quantum adder circuit as post quantum gate operation, given as in Fig.~\ref{nand_gate}. This circuit consists of three inputs: one ancillary input and two data inputs, where the ancillary input is $\ket{A}$ and the data inputs are $\ket{q_0}$ and $\ket{q_1}$. When the NAND gate is treated as a NOT-AND operation, its quantum counterpart can be realized by applying an additional X-gate (NOT) is applied to the $\ket{A}$ qubit. The output of the NAND gate is then obtained by measuring the $\ket{A}$ qubit, yielding the classical result $\mathrm{c} = q_0 \,\text{NAND}\, q_1$. 

This circuit was tested for all input states, $\ket{00}$,$\ket{01}$, $\ket{10}$ and $\ket{11}$, and the results were verified as given in Table~\ref{tab:nand_gate_measurements}
\begin{table}[ht]
\caption{\label{tab:nand_gate_measurements}Truth table of the QFT-based 2-input NAND-gate.}
\noindent
\begin{tabularx}{\columnwidth}{@{}XXX@{}}
\toprule
$\ket{q_0}$ & $\ket{q_1}$ & $c$ \\
\midrule
$\ket{0}$ & $\ket{0}$ & \text{1} \\
$\ket{0}$ & $\ket{1}$ & \text{1} \\
$\ket{1}$ & $\ket{0}$ & \text{1} \\
$\ket{1}$ & $\ket{1}$ & \text{0} \\
\bottomrule
\end{tabularx}
\end{table}

\subsection{\label{OR}Implementation of the logic OR gate}
\begin{figure}[!htb]\centering
\includegraphics[width=8.5cm]{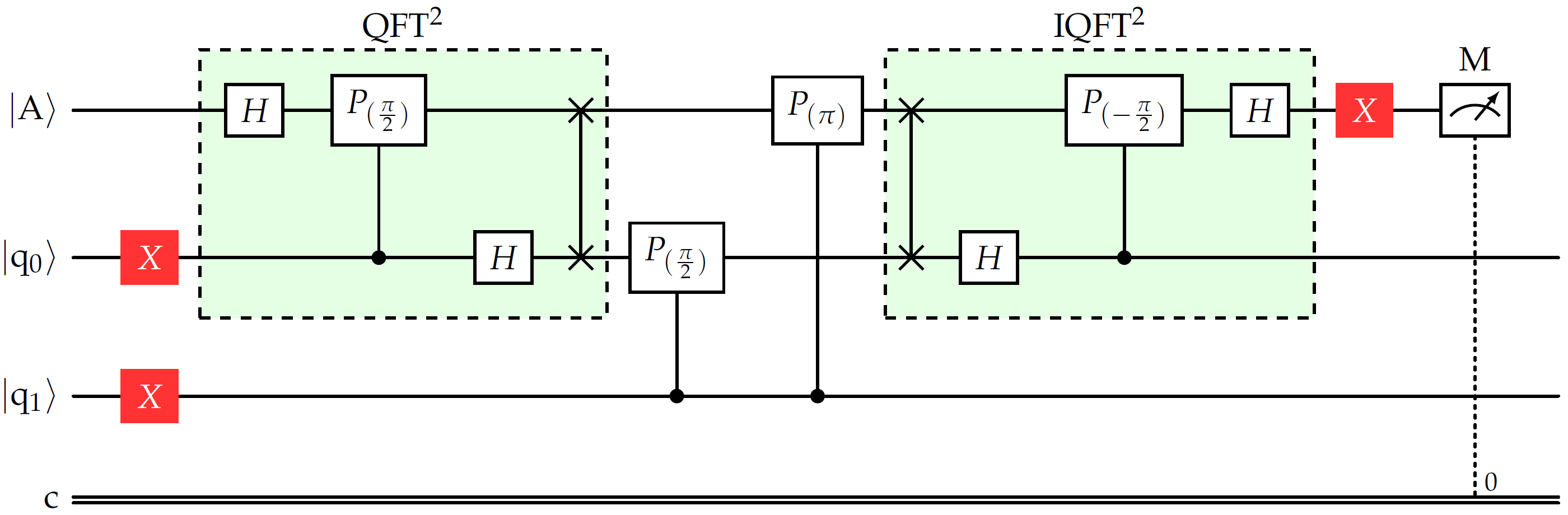}
\caption{\label{or_gate} The QFT-adder based two-input OR gate.}
\end{figure}

In this case, if the half-adder does not include an AND gate, it is not possible to implement the OR gate directly. So, we utilize De~Morgan's law of intersection, expressed as $\mathrm{\overline{X} \,\text{AND}\, \overline{Y} = \overline{X\, \text{OR}\, Y}}$, where $\overline{X}$ denotes the logical NOT of $X$. To implement the OR gate, pre- and post-processing quantum gates are applied to the one-bit, two-input QFT-based adder circuit. Specifically, NOT gates are first applied to both data inputs, $\ket{q_0}$ and $\ket{q_1}$, as a pre-processing step. Then, a NOT gate is applied to the ancillary qubit $\ket{A}$ of the QFT-adder as a post-processing step, in order to remove the negation in $\overline{q_0\, \text{OR}\, q_1}$, as illustrated in Fig.~\ref{or_gate}. The measurement outcome of the ancillary qubit corresponds to the result of the OR operation in the classical bit, $\mathrm{c} = q_0\,\text{OR}\, q_1$.

We test the all input states ($\ket{00}$, $\ket{01}$, $\ket{10}$ and $\ket{11}$) in IBM quantum composer. The results were verified as given in Table~\ref{tab:or_gate_measurements}.
\begin{table}[ht]
\caption{\label{tab:or_gate_measurements}Truth table of the QFT-based 2-input OR-gate. }
\noindent
\begin{tabularx}{\columnwidth}{@{}XXX@{}} 
\toprule
$\ket{q_0}$ & $\ket{q_1}$ & $c$ \\
\midrule
$\ket{0}$ & $\ket{0}$ & \text{0} \\
$\ket{0}$ & $\ket{1}$ & \text{1} \\
$\ket{1}$ & $\ket{0}$ & \text{1} \\
$\ket{1}$ & $\ket{1}$ & \text{1} \\
\bottomrule
\end{tabularx}
\end{table}

\subsection{\label{NOR}Implementation of the Logic NOR Gate Using a QFT-based circuit}
\begin{figure}[!htb]\centering
\includegraphics[width=8.5cm]{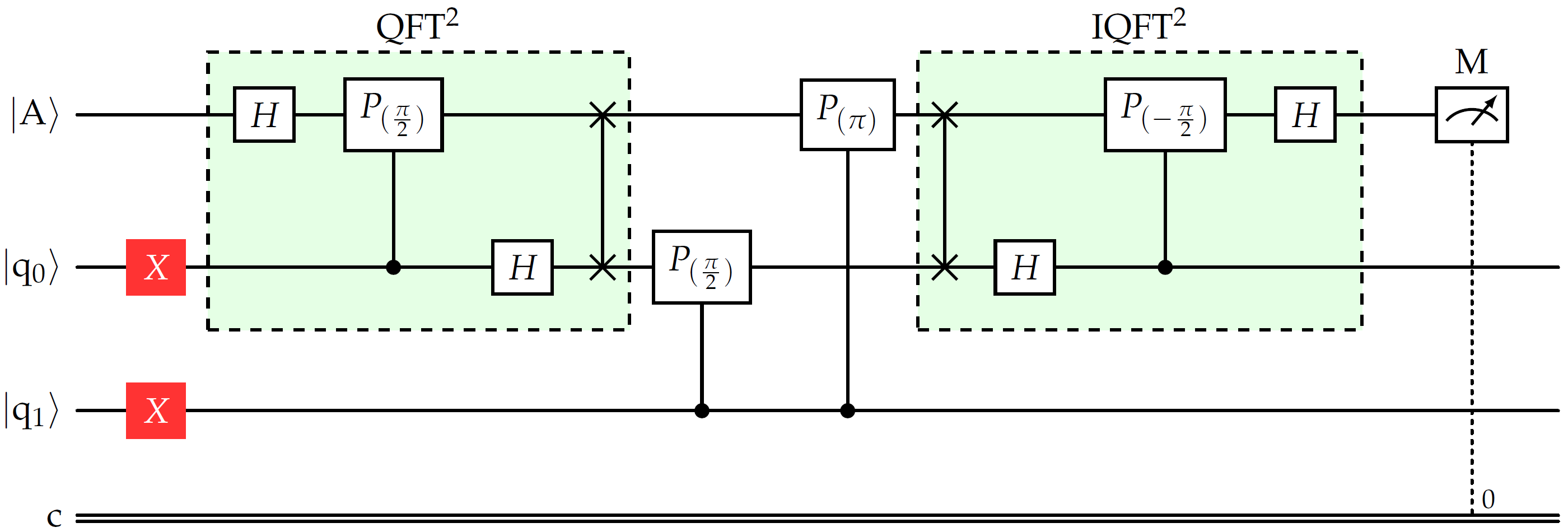}
\caption{\label{nor_gate} The QFT-adder based two-input NOR gate.}
\end{figure}

Similar to the OR gate, to design a two-input logic NOR gate, we apply pre-process to one-bit two-input QFT-based quantum adder circuit. The pre-process is realized two X-gate (NOT) to each data input of QFT-adder, $\ket{q_0}$ and $\ket{q_1}$. This provides the De Morgan's law which is $\mathrm{\overline{X} \,\text{AND}\, \overline{Y} = \overline{X\, \text{OR}\, Y}}$. The circuit is presented in Fig.~\ref{nor_gate} now operates NOR gate without applying post-process gate. The result of the NOR operation, can be measured from the ancillary qubit, is stored in classical bit, $\mathrm{c} = q_0 \,\text{NOR}\, q_1$.

The all input states ($\ket{00}$,$\ket{01}$, $\ket{10}$ and $\ket{11}$) were tested in IBM quantum composer and the results were verified that represented in Table \ref{tab:nor_gate_measurements}
\begin{table}[ht]
\caption{\label{tab:nor_gate_measurements}Truth table of the QFT-based 2-input NOR-gate.}
\noindent
\begin{tabularx}{1.0\columnwidth}{@{}XXX@{}} 
\toprule
$\ket{q_0}$ & $\ket{q_1}$ & $c$ \\
\midrule
$\ket{0}$ & $\ket{0}$ & \text{1} \\
$\ket{0}$ & $\ket{1}$ & \text{0} \\
$\ket{1}$ & $\ket{0}$ & \text{0} \\
$\ket{1}$ & $\ket{1}$ & \text{0} \\
\bottomrule
\end{tabularx}
\end{table}

\subsection{\label{XOR}Implementation of the logic XOR gate}
\begin{figure}[!htb]\centering
\includegraphics[width=8.5cm]{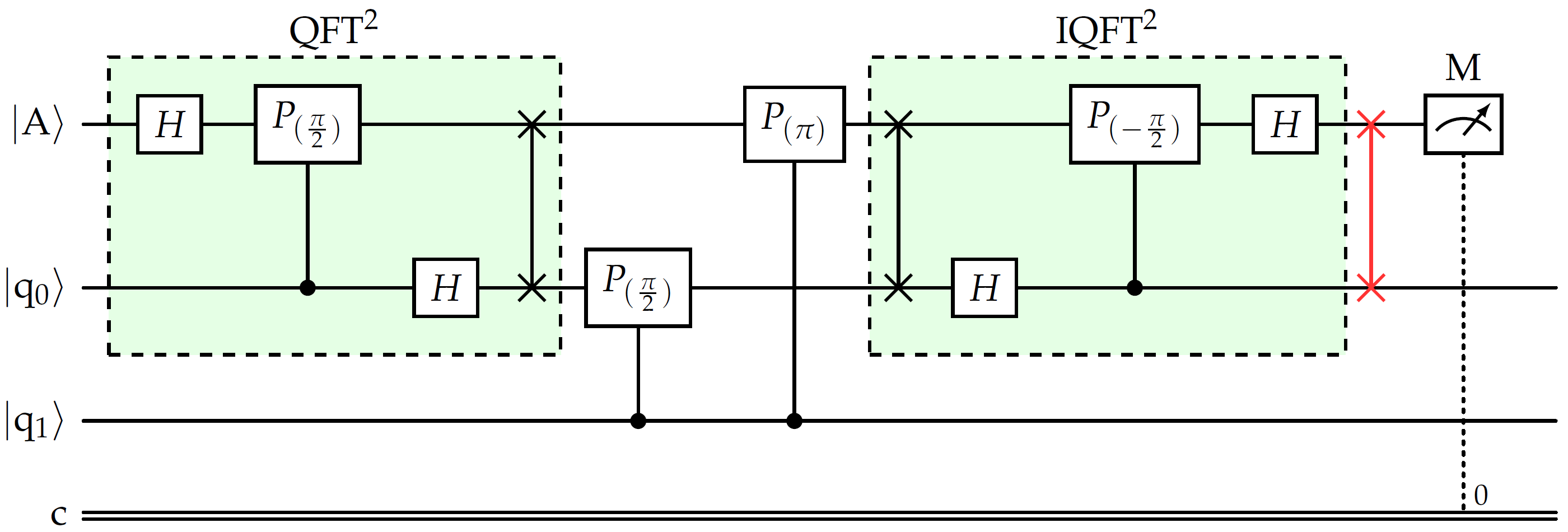}
\caption{\label{xor_gate} The QFT-adder based logic XOR gate.}
\end{figure}

In the final case, we construct the quantum counterpart of the XOR gate based on the one-bit, two-input QFT-adder circuit. Since the quantum adder replicates the behavior of a digital half-adder, the sum output of the QFT-adder corresponds to the XOR gate. In our generic design shown in Fig.~\ref{PrePostQFTadder}, the measurement is performed on the ancillary qubit $\ket{A}$. Therefore, we apply a SWAP gate as a post-processing step between $\ket{A}$ and $\ket{q_0}$ to transfer the XOR information, originally held by the $\ket{q_0}$ qubit, to the $\ket{A}$ qubit, as illustrated in Fig.~\ref{xor_gate}. As a result, the measurement outcome is stored in the classical bit, $\mathrm{c} = q_0 \,\text{XOR}\, q_1$. 
The Table~\ref{tab:xor_gate_measurements} verifies the test results of all input states, $\ket{00}$,$\ket{01}$, $\ket{10}$ and $\ket{11}$, on IBM quantum composer.
\begin{table}[ht] \centering
\caption{\label{tab:xor_gate_measurements}Truth table of the QFT-based XOR-gate.}
\noindent
\begin{tabularx}{1.0\columnwidth}{@{}XXX@{}} 
\toprule
$\ket{q_0}$ & $\ket{q_1}$ & $\mathrm{c}$ \\
\midrule
$\ket{0}$ & $\ket{0}$ & \text{0} \\
$\ket{0}$ & $\ket{1}$ & \text{1} \\
$\ket{1}$ & $\ket{0}$ & \text{1} \\
$\ket{1}$ & $\ket{1}$ & \text{0} \\
\bottomrule
\end{tabularx}
\end{table}

\section{Conclusion}
In this work, we construct a resource-efficient quantum architecture for the quantum counterparts of classical logic gates (AND, NAND, OR, NOR, and XOR) using a QFT-based adder. Our approach employs both pre-processing and post-processing quantum gates with the QFT-based adder to implement Boolean algebra properties, including De Morgan's laws. For each target logic operation acting on input qubits $\ket{q_0}$ and $\ket{q_1}$, we determine the output by measuring the ancillary qubit $\ket{A}$ and encoding the result into a classical bit channel.

While quantum counterparts of classical logic gates can indeed be implemented using only elementary quantum gates (without employing the quantum Fourier transform), such approaches typically exhibit an increasing qubit requirement as the input size scales~\cite{Cakmak24}. In contrast, the QFT-based adder architecture proposed in this work simultaneously reduces both the qubit count and improves noise resilience under quantum resource constraints. This approach may consequently enable more efficient realization of large-scale digital circuits in quantum computing systems.

\begin{backmatter}
\bmsection{Funding}
The authors acknowledge support from the Scientific and Technological Research Council of Turkey (T\"{U}B\.{I}TAK-Grant No. 122F298).

\bmsection{Disclosures}
The authors declare no conflicts of interest.

\end{backmatter}


\begin{thebibliography}{0}

\bibitem{Bennet2000} Bennett, C.H., DiVincenzo, D.P.: Quantum information and computation. Nature. 404, 247–255 (2000). 

\bibitem{Chae2024} Chae, E., Choi, J., Kim, J.: An elementary review on basic principles and developments of qubits for quantum computing. Nano Convergence. 11, 11 (2024).

\bibitem{Harrow17} Harrow, A.W., Montanaro, A.: Quantum computational supremacy. Nature. 549, 203–209 (2017). 

\bibitem{Boixo18} Boixo, S., Isakov, S.V., Smelyanskiy, V.N., Babbush, R., Ding, N., Jiang, Z., Bremner, M.J., Martinis, J.M., Neven, H.: Characterizing quantum supremacy in near-term devices. Nature Phys. 14, 595–600 (2018).

\bibitem{Arute19} Arute, F., Arya, K., Babbush, R., et al.: Quantum supremacy using a programmable superconducting processor. Nature. 574, 505–510 (2019). https://doi.org/10.1038/s41586-019-1666-5

\bibitem{Nielsen12} Nielsen, M.A., Chuang, I.L.: Quantum Computation and Quantum Information: 10th Anniversary Edition. Cambridge University Press (2012)

\bibitem{Shor97} Shor, P.W.: Polynomial-Time Algorithms for Prime Factorization and Discrete Logarithms on a Quantum Computer. SIAM J. Comput. 26, 1484–1509 (1997).

\bibitem{Grover96} Grover, L.K.: A fast quantum mechanical algorithm for database search. In: Proceedings of the twenty-eighth annual ACM symposium on Theory of computing  - STOC ’96. pp. 212–219. ACM Press, Philadelphia, Pennsylvania, United States (1996)

\bibitem{Camps21} Camps, D., Van Beeumen, R., Yang, C.: Quantum Fourier transform revisited. Numerical Linear Algebra App. 28, e2331 (2021).

\bibitem{Draper20} Draper, T.G.: Addition on a Quantum Computer, https://arxiv.org/abs/quant-ph/0008033, (2000)

\bibitem{Ruiz17} Ruiz-Perez, L., Garcia-Escartin, J.C.: Quantum arithmetic with the quantum Fourier transform. Quantum Inf Process. 16, 152 (2017). 

\bibitem{Sahin20} Şahin, E.: Quantum arithmetic operations based on quantum fourier transform on signed integers. Int. J. Quantum Inform. 18, 2050035 (2020).

\bibitem{Cakmak24} Çakmak, S., Kurt, M., Gençten, A.: Quantum Fourier Transform‐Based Arithmetic Logic Unit on a Quantum Processor. Annalen der Physik. 

\bibitem{Abdelgaber2020} Abdelgaber, N., Nikolopoulos, C.: Overview on Quantum Computing and its Applications in Artificial Intelligence. In: 2020 IEEE Third International Conference on Artificial Intelligence and Knowledge Engineering (AIKE). pp. 198–199. IEEE, Laguna Hills, CA, USA (2020)

\bibitem{Jozsa98} Jozsa, R.: Quantum algorithms and the Fourier transform. Proc. R. Soc. Lond. A. 454, 323–337 (1998). 

\bibitem{Barenco96} Barenco, A., Ekert, A., Suominen, K.-A., Törmä, P.: Approximate quantum Fourier transform and decoherence. Phys. Rev. A. 54, 139–146 (1996).

\bibitem{Cao11} Cao, Y., Peng, S.-G., Zheng, C., Long, G.-L.: Quantum Fourier Transform and Phase Estimation in Qudit System. Commun. Theor. Phys. 55, 790–794 (2011).

\bibitem{Pavlidis17} Pavlidis, A., Floratos, E.: Quantum-Fourier-transform-based quantum arithmetic with qudits. Phys. Rev. A. 103, 032417 (2021).

\bibitem{Kurt24} Kurt, M., Kaltehei, A., Gençten, A., Çakmak, S.: Scalable quantum circuit design for QFT-based arithmetic, https://arxiv.org/abs/2411.00260, (2024)

\bibitem{IBMqc} IBM Quantum. IBM Quantum Composer. Available at: \url{https://quantum.ibm.com/composer}. Accessed April 23, 2025

\bibitem{Renaud11} Renaud, N., Joachim, C.: Classical Boolean logic gates with quantum systems. J. Phys. A: Math. Theor. 44, 155302 (2011). 

\bibitem{Paul23} Paul, B., Choudhury, N., Saikia, E., Trivedi, G.: Digital Boolean Logic Equivalent Reversible Quantum Gates Design. In: Kumar, S., Sharma, H., Balachandran, K., Kim, J.H., and Bansal, J.C. (eds.) Third Congress on Intelligent Systems. pp. 253–271. Springer Nature Singapore, Singapore (2023)

\bibitem{Miller11} Miller, D.M., Wille, R., Sasanian, Z.: Elementary Quantum Gate Realizations for Multiple-Control Toffoli Gates. In: Proc. 41st IEEE International Symposium on Multiple-Valued Logic, pp. 288–293. IEEE (2011)

\end{thebibliography}
\end{document}